\documentclass[
twocolumn,aps,prl
]{revtex4-2}

\usepackage[english]{babel}

\usepackage{graphicx,color,hyperref}

\usepackage{amsmath}
\usepackage{amsfonts}
\usepackage{amssymb}
\usepackage{dsfont}
\usepackage{braket}

\newcommand{\tr}{\mathrm{tr}}

\begin{document}

\title{
Tensor product random matrix theory 
}

\author{Alexander Altland}
\affiliation{Institut f\"ur Theoretische Physik, Universit\"at zu K\"oln, Z\"ulpicher Str. 77, 50937 Cologne, Germany
}

\author{ 
Joaquim Telles de Miranda
}
\affiliation{
Centro Brasileiro de Pesquisas F\'isicas, Rua Xavier Sigaud 150, 22290-180, Rio de Janeiro, Brazil 
} 

\author{ 
Tobias Micklitz
}
\affiliation{
Centro Brasileiro de Pesquisas F\'isicas, Rua Xavier Sigaud 150, 22290-180, Rio de Janeiro, Brazil 
} 

\date{\today}


\begin{abstract}
The evolution of complex correlated quantum systems such as  random circuit
networks is governed by the dynamical buildup of both entanglement and entropy.
We here introduce a  real-time field theory approach --- essentially a fusion of
the $G \Sigma$ functional of the SYK model and the field theory of disordered
systems ---  enigneered to microscopically describe the full range of such crossover
dynamics: from initial product states to a  maximum entropy
ergodic state. To showcase this approach in the simplest nontrivial setting, we
consider a tensor product of coupled random matrices, and compare to exact
diagonalization. 
\end{abstract}

\maketitle

{\it Introduction:---}When addressing the physics of complex many-body systems,
an early choice needs to be made: One option is to focus on  individually
resolved quasiparticle trajectories, which in concrete terms means working with
second quantized representations, many-body field integrals, and related
concepts. The alternative is to consider the more complex trajectory of the system at
large, as described by a single exponentially high-dimensional Hamiltonian matrix. Depending on the context, the preferred choice
is usually evident, use cases including large portfolio of many-body field
theory approaches to condensed matter systems, or the modeling of
complex nuclei in terms of high dimensional random matrix theories, respectively. However, since the advent of synthetic quantum matter, we are
increasingly witnessing applications in twilight zones, where the appropriate
approach is not so clear. For example, starting from initial product states of
qudits or quasiparticles, circuit dynamics addresses the buildup of
complexity,  eventually leading to a fully entangled state. In the latter,
individual states can no longer be meaningfully resolved and global
representations, e.g., in terms of a high-dimensional random matrices, may be more appropriate.  
The crux of the matter is that precisely the crossover dynamics leading from
individually resolved quasiparticle world lines to the single complex trajectory
of a many-body state is hard to capture in terms of the analytical toolkits of
field or path integration, which may be a reason for the relative scarcity of
such methods in the theory of quantum circuits. (For powerful phenomenological,
semiclassical, or statistical mechanics inspired approaches, see, e.g.,
Refs.~\cite{chanSpectralStatisticsSpatially2018,nahumQuantumEntanglementGrowth2017,vasseurEntanglementTransitionsHolographic2019,royPercolationFockSpace2019,PhysRevX.8.041019,PhysRevX.11.021051,liaoUniversalDephasingMechanism2022}.)

In this paper we define a  
random matrix model displaying such type of crossover dynamics in the simplest
possible setting. The model will serve as a test bed for a path integral
approach whose central ingredient  is an interaction vertex possessing the
mathematical structure of a Hadamard product. We will discuss how the Hadamard
vertex catalyzes the progressive locking of initially separate subsystem
trajectories to a configuration representing a maximally entangled ergodic final
state. As a representative of various observables witnessing this dynamics we
will consider the spectral form factor, whose time dependence we will compute
analytically and compare to exact diagonalization. We conclude with a discussion
of the extensibility of the approach exemplified here to more complex circuit
arrays.

\begin{figure}[t!]
    \centering
    \includegraphics[width=.8\columnwidth]{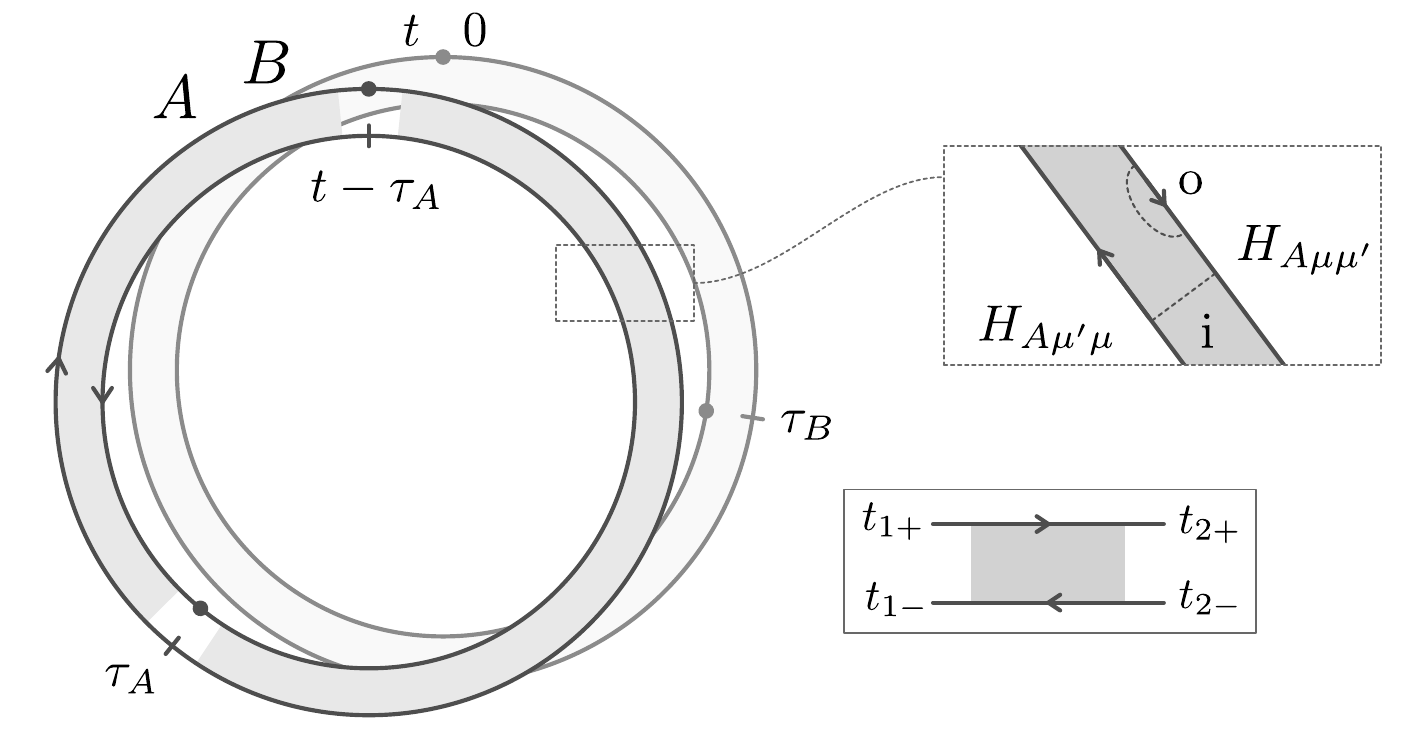}
    \caption{ Schematic representation of closed scattering sequences in $A$
    and $B$ Hilbert space traversed in retarded and advanced time order (outer and inner
    rings), and starting and ending at times 0 and $t$, respectively. 
    For constructive interference, the sequences must contain
    the same scatterers but can be traversed at a relative time delay, $\tau_A$
    and $\tau_B$, indicated by dots on the retarded and advanced contours.
    Insets: In- and out-scattering induced by $H_A$, and temporal structure of
    the Gaussian propagator Eq.~\eqref{eq:Propagator}.} 
    \label{fig1}
\end{figure}

The model comprises two $N_X$-dimensional qudits, $X=A,B$, governed by the
Hamiltonians $H_A \otimes \mathds{1}_B$ and $\mathds{1}_A \otimes H_B$, with
$H_X$  drawn from a Gaussian random matrix ensemble. The two sectors are
entangled via the likewise random $N_A N_B$-dimensional Hamiltonian $H_{AB}$,
defining the full Hamiltonian $H=H_A + H_B + H_{AB}$, 
where we suppress trivial factors $\otimes \mathds{1}$ for notational simplicity.  
Floquet variants of this model have been recently 
 studied with a focus on  spectra~\cite{Kieler2024,Backer2016}, 
 eigenstates~\cite{Tomsovic2016}, and
entanglement properties~\cite{Tomsovic2020,Tomsovic2016,Backer2018,Backer2021}. While these analyses were based on combinations of  semiclassical, perturbative, and numerical approaches, we here aim for a first-principles analysis of the Hamiltonian system by field integral methods. 
 
We begin our analysis of this  model  with a
qualitative discussion in terms of Feynman path amplitudes. This picture  will
provide an intuitive basis for the subsequent quantitative formulation in terms
of a real-time matrix path integral.

{\it Qualitative discussion:} In the absence of coupling, the spectrum of our
model comprises all $N_A N_B$ different sums $\epsilon_{A i}+ \epsilon_{Bj}$.
While these levels can come arbitrarily close (no RMT-like level repulsion) they
remain statistically correlated (no Poisson statistics). To understand these
correlations, it is best to switch to time space, and consider the spectral form
factor, which, prior to normalization, is defined as $K(t)=\langle
\tr(G^+(t))\tr(G^-(-t)) \rangle$ in terms of the retarded and advanced Green's
functions $G^\pm(t)=\mp i \Theta\left(\pm t\right)\exp(-i H t)$. Consider first
a single ergodic system, such as the $A$ factor of our tensor product structure.
Representing the trace as $\tr(\dots)=\sum_\mu \langle \mu | \dots | \mu
\rangle$, the form factor becomes a double sum over self-returning scattering
paths traversed in time $t$, as indicated in Fig.~\ref{fig1}. To leading
approximation in $N_A^{-1}$, only paths traversing the same sequence of
scattering events in retarded and advanced chronology contribute upon averaging
over the ensemble. However, the anchor points at which the sequences start in
Hilbert space at  time $t=0$ can be independently chosen for the two amplitudes,
and it is this sliding degree of freedom which is responsible for the
linearity $K(t)\sim t$~\cite{Haake}.  Turning to the full problem, the trace
assumes the form $\tr(\dots)=\sum_{\mu \nu }\langle \mu \nu| \dots |\mu \nu
\rangle$, where we use the shorthand notation $\langle \mu \nu |\equiv \langle
\mu|\otimes \langle \nu|$, and the second factor labels the states of $B$. On
this basis, we now compare the two alternative approaches mentioned above, paraphrased as second vs. first
quantized for brevity. 

{\it Second quantized:} Consider the states $|\mu \nu\rangle = a_\mu^\dagger
b_\nu^\dagger|0\rangle $ represented in terms of single-particle bosonic
 operators generating the subsystem states. Beginning with the uncoupled
system, $H_{AB}=0$, the factorization of the trace
$\tr(G^\pm(t))=\tr(G^\pm_A(t))\tr(G^\pm_B(t))$ is manifest in this language. It
extends to the form factor, and so $K(t)\sim t^2$ in this limit. However, upon
switching on the interaction, we encounter the \emph{nonlinear} operator
(summation convention) $H_{AB}= a_\mu^\dagger b_\nu^\dagger H_{\mu \nu,\mu'
\nu'}a_{\mu'}b_{\nu'}$, expected to   send the full system into an ergodic
regime. Defining the variances of our random Hamiltonians as $N_A\langle
|H_{A\mu \mu'}|^2 \rangle= N_B \langle |H_{B\nu \nu'}|^2 \rangle= \lambda^2$,
and $N_A N_B\langle |H_{AB\mu \nu,\mu' \nu' }|^2 \rangle= \Lambda^2$, a
straightforward Golden Rule estimate suggests a coupling rate 
$\gamma \propto\Lambda^2 /\lambda$, 
and hence we expect $K(t)\sim t $ for time scales $t>
\gamma^{-1}$.       However, the quantitative description of this crossover
requires the solution of an interacting problem, which is nontrivial in this
language. 

{\it First quantized:} In this representation, $H_{\mu \nu,\mu' \nu'}$ is
a matrix of dimension $N_A N_B$, and the derivation  of a linear form factor $K(t)\sim t$
follows standard protocols (for example, by interpreting the construction
indicated in Fig.~\ref{fig1} in terms of scattering paths in tensor product
space.) However, it is no longer straightforward to describe the crossover to
$K(t)\sim t^2$ upon \emph{lowering} $\Lambda$; extracting the disentangling into separate subsystem paths in $\mu$ and $\nu$ space and the
emergence of two independent sliding symmetries from the global matrix representation is difficult. We observe that  neither approach appears to be well suited to address the middle ground between the two limiting regimes.

{\it Field integral approach:} In the following, we follow a  strategy
inspired in equal parts by  field theories of disordered systems~\cite{Efetbook},
and the $G \Sigma$ approach to the SYK model (for a review, see
Ref.~\cite{rosenhausIntroductionSYKModel2019}), to construct a theory
tailored to describe the form factor at all time scales.  Referring to the Supplemental Material for details~\cite{sup_mat}, 
we start in second quantization and next turn to  a real-time field integral
representation, $K(t)=\langle \prod_{s=+,-}a^s_{\mu,t} b^s_{\nu,t}\bar
a^s_{\mu,0} \bar b^s_{\nu,0} \rangle$, where $a=\{a_{\mu,t}^s\} $ and
$b=\{b_{\nu,t}^s\}$ are now complex commuting field variables, and the causal
index ($s=\pm$) distinguishes between fields propagating forward and backward in
time. The functional average $ \langle \dots \rangle=\int D(a,b)\exp(i
S[a,b])(\dots)$ is over the  action $S[a,b]=S_0[a,b]+ S_\textrm{int}[a,b]$, with
$S_0[a,b]=\int_0^t dt \sum_{x=a,b}\bar x \tau_3(i \partial_t - H_X+i 0 \tau_3
)x$, and $S_\textrm{int}[a,b]=\int dt\, \bar a \bar b\,  \tau_3H_{AB} a b$, where
Hilbert space indices are suppressed for clarity, and $\tau_3$ is a Pauli matrix
in contour space. The (It\^o) time discretization of the functional integral implies
$\langle 1 \rangle=1$, i.e. it yields real-time Green's functions without the need
for explicit normalization~\cite{Kamenev_2011}. 

Averaging over the distribution of the three involved random matrices produces
an effective action containing products of up to eight field operators. They all
occur in combinations summed over Hilbert space indices, suggesting the
introduction of effective Hubbard-Stratonovich variables $(G_{X})_{t t'}^{s
s'}\equiv - i N_X^{-1} \, x_{\mu t}^s \bar x_{\mu t'}^{s'}(-)^{s'}$. We effect
this locking via Lagrange multipliers $(\Sigma_X)_{t t'}^{s s'}$, and arrive at
the representation $\langle \dots \rangle=\int D(G, \Sigma)\,\exp(i
S[G,\Sigma])$, where $S=S_A+S_B + S_{AB}$, with 
\begin{align}
    \label{eq:GSigmaAction}
    &S_X[G,\Sigma]=   iN_X \tr\left(\ln (i \partial_t  - \Sigma_X) + G_X \Sigma_X + \frac{\lambda^2}{2} G_X^2\right),\nonumber\\  
    &S_{AB}[G]=  i \frac{\Lambda^2 N_A N_B }{2}\tr ((G_A \odot G_B \tau_3)^2).
\end{align}
Here, the traces imply summation over all indices, $\tr(AB)=\int dt dt' A^{s
s'}_{t t'}B^{s' s}_{t't}$, and we encounter the Hadamard or element-wise
product  defined as $(A\odot B)^{s s'}_{t t'}= A^{s s'}_{t t'}B^{s s'}_{t t'}$. In the same notation, the correlation function assumes the form $K(t)= (N_A N_B)^2\langle (G_A \odot G_B)^{++}_{t,0} (G_A \odot G_B)^{--}_{0,t}
\rangle$.

{\it Stationary phase:} The introduction of the collective $(G,\Sigma)$ variables is rewarded by the appearance of
the factors $N_X$ upfront the action, inviting a stationary phase approach.   
For simplicity, we assume that the  coupling $H_{AB}$ is sufficiently weak to
not significantly affect the mean-field Green's functions of our system.
Temporarily neglecting the Hadamard vertex, a variation of the action in  $G$
and $\Sigma$ (we  suppress the label $X=A,B$ for readability) then leads to the
equations $G = (i \partial_t - \Sigma )^{-1} $ and $\Sigma= \lambda^2 G$, which
are solved by $\bar G_{t t'}= -i \tau_3 \Theta(\tau_3(t-t'))\exp(-\lambda
|t-t'|)$: the average Green's functions and the self-energies are rapidly decaying
functions in time, respecting causality. Neglecting $\partial_t$ in comparison
to $\Sigma$, the saddle point equations are approximately invariant under
temporally slow rotations, $\bar G \to T \bar G T^{-1}$,  $T=\{T_{t t'}^{s s'}\}
$. (Note that the rotational symmetry breaking in causal space, $\bar G \propto
\tau_3$, identifies the $T$'s as Goldstone mode fluctuations, with $\partial_t$
an `explicit symmetry breaking'.) The expansion of the action to leading
order in derivatives $\partial_t$ reads as $S[T_A,T_B]=\sum_X S_{\rho_X}[T_X]$,
where 
\begin{align}
    \label{eq:GoldstoneModeAction}
    S_\rho[T]= i\pi\rho \int  dt dt' \tr(\tau_3 T_{tt'}^{-1}(\partial_{t'}+\partial_t)T_{t't}),    
\end{align}
 is the real-time representation of the matrix actions known to produce the
random matrix form factor of systems with spectral density
$\rho_X=N_X/(\pi\lambda)$~\cite{Efetbook,Kamenev1999}. With the notation
$Q=T \tau_3 T^{-1}$, our correlation function  assumes the form $K(t)=\rho_A^2
\rho_B^2 \prod_X\langle Q^{++}_{Xt,0}Q^{--}_{X0,t}  \rangle$, the
factorization implying that  at this level we obtain the product form factor of
the decoupled theory.

However, what is not so evident is how for finite $\Lambda$ the Hadamard
coupling vertex will result in a collapse to a single Goldstone mode $Q$ required
to describe an emerging ergodic phase. Naively, one might speculate that its
proportionality to $G_A \odot G_B \sim Q_A\odot Q_B$ enforces a locking $Q_A =
Q_B =Q$ of the fluctuation fields. However, this is not how it works. It is
straightforward to verify that the vertex does not vanish on such field
configurations, implying that sought-after modes must be realized differently.
The crossover problem articulated in the beginning now assumes a very concrete
form: our path integral must possess a single Goldstone mode with a uniquely
specified action. However, it is not obvious where this mode hides.

\begin{figure}[t!]
    \centering
    \includegraphics[width=.9\columnwidth]{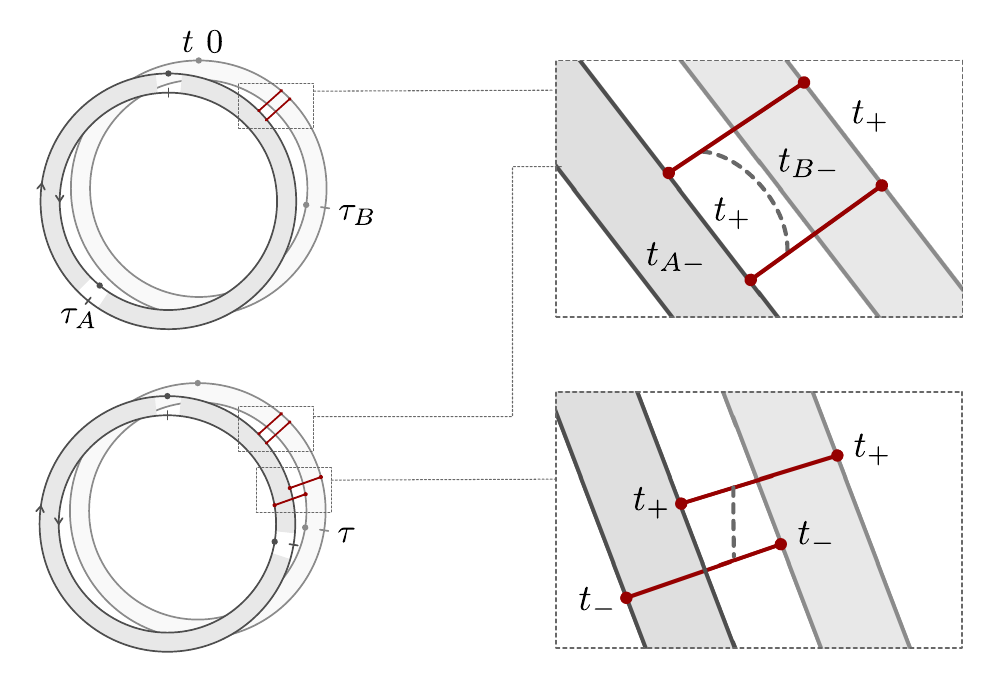}
    \caption{ Correlation function in the presence of interactions. Top: For
    scattering path traversing Fock space in generic, asynchronous order, only
    the two-body generalization of out-scattering or self-energy damping is
    operational. Bottom: for time-synchronized path, or equal-time scattering
    sequences in tensor space, in-scattering cancels against out-scattering,
    stabilizing an undamped ergodic model.} 
    \label{fig2}
\end{figure}

{\it Perturbation theory: }In order to get traction with this problem, and
establish contact to the above Feynman path picture, we turn to a more concrete
level and expand the Goldstone mode fluctuations in generators chosen to
anti-commute with the saddle point $\sim\tau_3$,  $T=\exp\left(B \tau_+ -
B^\dagger \tau_-\right)$, where $\tau^\pm = \frac{1}{2}(\tau_1 \pm i \tau_2)$
and  $B=\{B_{t t'}\}$\cite{Kamenev1999}. The quadratic expansion of the
decoupled theories assumes the form $S[B]= 2i\pi \rho \int dt dt' B_{t
t'}(\partial_t + \partial_{t'})B^\dagger_{t't}$, implying that the Wick
contraction of matrices in this free quadratic theory is given by 
\begin{align}
    \label{eq:Propagator}
    \langle  B_{t_{1+} t_{1-} }B^\dagger_{t_{2-}t_{2+}} \rangle= \Delta \delta(\Delta t_1-\Delta t_2)\Theta (t_2 - t_1),
\end{align}           
with $\Delta=1/(2\pi\rho)$, $\Delta t = t_+ - t_-$ and $ t=(t_+ + t_-)/ 2$, and
the $\delta$ function broadened on the minimal timescale resolved by the
$B$ theory, $\lambda^{-1}$. Physically, these modes describe the  pair
propagation of states  indicated in the lower inset of Fig.~\ref{fig1}, where
the temporal constraints imply equal propagation time of the retarded and
advanced amplitude, $t_{2+} - t_{1+}=t_{2-} - t_{1-}$, and causality
$t_{2s}>t_{1s}$. Otherwise, the $B$ propagator is structureless, reflecting
ergodicity.  Substitution of the expansion $G^{++}_{t,0}\to \textrm{const.}+
\int d \tau B_{t,\tau}B^\dagger_{\tau,0}$ and application of the contraction
yields each of the correlation functions as a product of two modes
(Fig.~\ref{fig1}), leading to $K(t) = \prod_X \left(\rho_X^2 \int_0^t d\tau_X
\cdot 1 \right)\propto t^2$, i.e. the product of two form factors, individually
proportional to $t$. The influence of the Hadamard vertex becomes transparent
once we subject it, too, to the generator expansion. Referring to the
Supplemental Material~\cite{sup_mat}, the latter starts at fourth order,
$S_{AB}=S_{\textrm{o}+}+S_{\textrm{o}-}+ S_\textrm{i}$ with
\begin{align}
    \label{eq:HadamardQuartic}
   S_{\textrm{o}+}[B]&= c_\textrm{o} \int dt_+ \prod_X \int dt_X \, 
   B_{X t_+ t_{X-}} B_{X t_{X-} t_+ }^\dagger,\\ 
S_\textrm{i}[B]&
   =- c_\textrm{i}\int dt_+ dt_- B_{A t_+ t_-} B_{B t_+ t_-} B^\dagger_{At_- t_+} 
   B^\dagger_{Bt_- t_+},\nonumber
\end{align}
indicating that (a) we are expanding around a proper saddle point and (b) it
does not induce a trivial mass as a quadratic term would. Here,
$c_\textrm{o}=2iN_A N_B\Lambda^2/\lambda^3 $, $c_\textrm{i}=16iN_A
N_B\Lambda^2/\lambda^4$, and the action $S_{\textrm{o}-}$ is obtained from
$S_{\textrm{o}+}$ by an exchange $B \leftrightarrow B^\dagger$.  
These expressions describe the two-body analogs of the elementary one-body
self-energy and vertex, \textrm{o} and $\textrm{i}$ indicated in the inset of
Fig.~\ref{fig1}. Specifically, $S_{\textrm{o},\pm}$ represents a damping term
correlating the previously independent pair amplitudes in $A$ and $B$ Hilbert
space through `out-' scattering processes, cf.  Fig.~\ref{fig2} top (for the
discussion of a similar damping mechanism for SYK-like models or coupled circuits, cf.
Refs.\cite{PhysRevX.11.021051,chanSpectralStatisticsSpatially2018,liaoUniversalDephasingMechanism2022}).
Coupling only to retarded  $(A+,B+)$, respectively, advanced $(A-,B-)$
amplitudes, it leaves the time differences $\tau_A=t_{A+}-t_{A-}$ and
$\tau_B=t_{B+}-t_{B-}$ unconstrained. By contrast, the competing (note the
opposite sign) `in-' scattering vertex does couple advanced and retarded
amplitudes, where the instantaneous of the scattering between the $A$ and $B$
sector requires $\tau_A=\tau_B$, cf. Fig.~\ref{fig2}, bottom.

Including these vertices into the computation of the correlation function, we
obtain the situation depicted in Fig.~\ref{fig2}. Prior to interactions (cf.
Fig.~\ref{fig1}), the correlation function $K(t)$ implies the independent
integration over two time arguments $\tau_X$, $X=A,B$, parametrizing the
temporal delay between retarded and advanced scattering paths. Upon switching on
interactions, generic of these amplitudes suffer out-scattering (top figure),
and hence get exponentially damped. However, for synchronous paths $\tau_A
\approx \tau_B\equiv \tau$ in-scattering balances the out-scattering and an
undamped contribution ensues (bottom figure). The integral over the single
parameter $\tau$ then produces the proportionality $K(t)\sim t$ indicative of
ergodicity. The projection onto synchronous paths implies that we are effectively
summing over trajectories in tensor product space $t\mapsto \ket{\mu \nu}(t)$,
rather than over products of single-particle trajectories $t\mapsto
\ket{\mu(t)}$ and $t\mapsto \ket{\nu(t)}$. 

A straightforward summation over interaction vertex insertions of arbitrary
order leads to the result 
\begin{align}
\label{eq:formfactor}
    K(t)=\frac{c}{N_AN_B} \left(
        (t\delta)^2  e^{- 2\gamma t
        }+t \delta 
\right),
\end{align} 
where $\gamma\approx \Lambda^2/(2\lambda)$, the coupling rate,
 $\delta\sim\lambda$ is set by the width of the spectrum, and $c$ a numerical
 factor of order ${\cal O}(1)$.  
The first and second term in Eq.~\eqref{eq:formfactor} are the exponentially
damped contribution of generic asynchronous pairs of single-qudit paths, and that of 
undamped tensor product trajectories, respectively. 
Our findings are in agreement with the recent results~\cite{Kieler2024} 
for the spectral form factor in a related Floquet model, and the chaotic 
regime of a complex SYK model~\cite{liaoUniversalDephasingMechanism2022}.

\begin{figure}[t!]
    \centering
    \vspace{-.4cm}
    \includegraphics[width=9cm,angle=0]{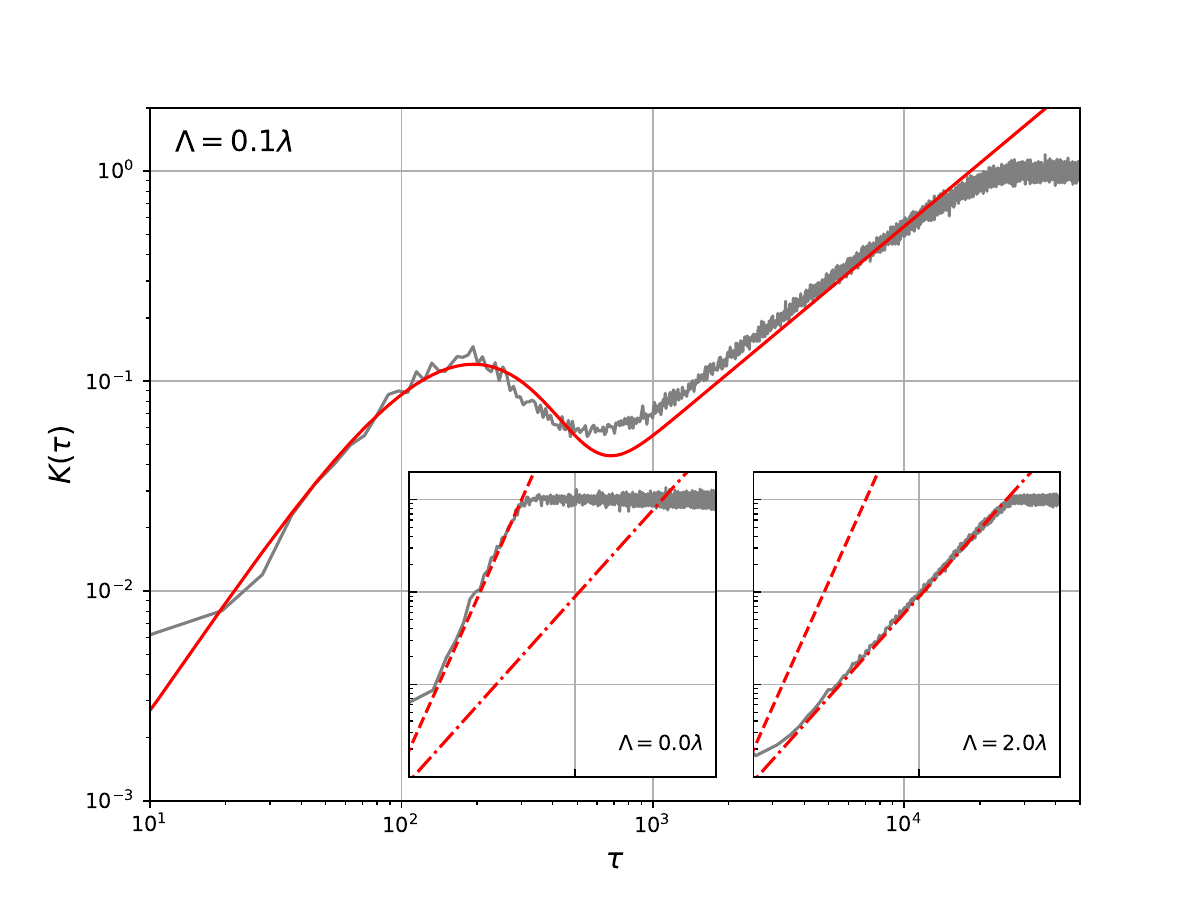}
    \caption{ 
    Comparison of the spectral form factor Eq.~\eqref{eq:formfactor} (red) and numerical simulations (gray), 
    for $\Lambda/\lambda=0,0.1$ and $2$. Dashed and dash-dotted lines in the insets indicate quadratic, respectively, 
    linear profiles. 
    For the main plot we use three fitting parameters, resulting in good agreement 
    (see main text for further discussion).
    }
    \label{fig3}
\end{figure}

{\it Numerical validation:} Before discussing how the analysis above can be
 pushed beyond the level of perturbation theory, let us discuss how it fares in
 comparison to numerics. Fig.~\ref{fig3} shows the analytical prediction
 Eq.~\eqref{eq:formfactor} and exact diagonalization results obtained by
 sampling over 500 realizations of the matrix model with dimensions $N_A = N_B =
 130$. The comparison involves three fit parameters, $(a,b,c)=(1.08,0.44,2.11)$
 defined through the proportionalities, $\gamma = a \Lambda^2/2\lambda$, $\delta
 = b \lambda$, and the normalization in Eq.~\eqref{eq:formfactor}, respectively.
 The reason why the values $a,b$  are left unspecified is that our definition of
 the form factor effectively integrates over all states in the spectrum, whereas
 the analytical calculation is based on approximations valid in the band center.
 While we have not been ambitious to push the computation to the higher level of
 refinement required for a parameter-free comparison, the nonmonotonous profile
 of the form factor predicted by Eq.~\eqref{eq:formfactor} at intermediate
 values of the coupling is quantitatively confirmed~\cite{footnote1}.

{\it Generalizations:} It is straightforward to generalize the above
construction beyond  perturbation theory, and in this way identify
the Goldstone mode action governing the system in the long time limit: Starting from 
the matrix representation of  observables as 
$\psi_{A t} \psi_{Bt}\bar \psi_{A t'} \bar\psi_{Bt'}\dots 
\to (Q_A \tau_3\odot Q_B \tau_3)_{t t'} \dots$, we imagine each of the constituent Goldstone modes expanded in generators $W=B \tau_+ - B^\dagger
\tau_-$, leading to expressions like 
$(W_A^n\odot W_B^n)_{tt'}$  
at $n$th order in the expansion. We now observe that each individual Wick
contraction applied in the reduction of such expansions must be subject to the
equal-time difference condition, $W_{A t,t+\tau_A} W_{B t,t+\tau_B}\to W_{A
t,t+\tau} W_{B t,t+\tau}=(W_A \odot W_B)_{t,t+\tau}$, to escape the
damping by out-terms. The recursive application of this condition, implies the collapse
$(W_A^n \odot W_B^n)_{t t'}\to ((W_A\odot W_B)^n )_{t t'}$. 
Finally, the individual generators appearing in this Hadamard product
must be contracted subject to the rule Eq.~\eqref{eq:Propagator}. With
Eq.~\eqref{eq:Propagator}, and the idempotency $\Theta^2(t-t')=\Theta(t-t')$, we
find that the product of $A$ and $B$ contractions is equivalent to a
single one,  $W_A\odot W_B \to W$, governed by the action
Eq.~\eqref{eq:GoldstoneModeAction}, where $Q=T \tau_3 T^{-1}$ and $T=\exp(W)$,
with the two-body density of states
$\rho \propto N_A N_B$. In this way, we
demonstrate how at time scales exceeding the coupling time, $t\sim  \lambda
/\Lambda^2$ the effective theory of an ergodic regime in a Hilbert space of
dimension  $N_A N_B$ emerges.        

{\it Summary and discussion:} We have constructed a path integral  describing the evolution of a  
tensor product of two weakly coupled random qudits. For times shorter than a
Golden Rule  coupling time, the latter evolve as nearly independent, but
individually ergodic subsystems, subject to weak damping due to the interqudit
relaxation. However, a crucial feature of the integral is that it selects
temporally synchronous paths which are exempt of damping and eventually combine
to define the ergodic phase of the coupled system. Alongside the semiclassical
picture, these contributions afford a nonperturbative description in terms of a Goldstone
mode, $Q$,  representing the universal causal symmetry breaking principle
indicative of ergodic quantum chaos.  Turning back to the tension between first
and second quantization raised in the beginning, we note that had we approached
the problem in  the former that Goldstone mode would have been visible from the
beginning~\cite{Efetbook}. However, in this formulation the opposite short time regime, requiring the emergence of two independent modes representing the subsystems looks inaccessible (or at least we have not managed to describe it within this framework.)

The exploration of a minimal problem prompts considerations regarding the
extensibility to more intricate cases, such as systems in the SYK class,
where the challenge of bridging between short time quasiparticle evolution~\cite{rosenhausIntroductionSYKModel2019} and
long time ergodicity~\cite{altlandQuantumErgodicitySYK2018} remains unresolved, or networks of coupled circuits.
While these  must be considered on a case-by-case basis, a few overarching
observations can be made. First, interacting models subject to statistically
independent random coefficients generically lead to path integrals with Hadamard
vertices. (A case in point is the $G \Sigma$-action of the  SYK model, which is
governed by an operator $G^{\odot 4}$ in the notation of the present paper.)
Second, we have seen that the emergence of a universal ergodic Goldstone mode
via this operator followed from a cancellation of self-energy (out) and vertex
(in) scattering processes, which in turn is a robust feature of unitarity. On
this basis, we are optimistic that the concepts introduced here, will become
instrumental in the analytical description of more complex use cases of
entanglement dynamics.

{\it Acknowledgements:---} A.A.  acknowledges partial support from the Deutsche
Forschungsgemeinschaft (DFG) through the Cluster of
Excellence Matter and Light for Quantum Computing (ML4Q) EXC 2004/1
390534769 and the CRC network TR 183 (project grant 277101999) as part of project A03. TM
acknowledges financial support by Brazilian agencies CNPq and FAPERJ, and J.T.M.
financial support by Brazilian agency CAPES.
{\bf Data and materials availability:} Processed data and python script used to generate 
Fig.~\ref{fig3} are available in Zenodo with identifier 10.5281/zenodo.10949792 \cite{zenodo}.

%



\section{Supplementary Material}

In this supplemental material we provide details
of the real time path integral construction for the tensor product random matrix model.

\subsection{Real time field integral}

We here provide some details of the real time path integral construction of our
tensor model. Starting from the second quantized representation of the trace
over the time evolution operator, $\tr(G^\pm(t))=\bra{0} a_\mu b^\nu \exp(\pm i
H t-0 t) a_\mu^\dagger b_\nu^\dagger\ket{0}$, trotterization in terms of a
standard bosonic coherent state representation~\cite{Altland2023}, leads to the
representation of the correlation function detailed in the main text, where the
causal index $s$ distinguishes between the two contour orientations. (For
comparison, note that we might just as well have worked with an algebra of
creation operators $\{X_{\mu \nu}\}$ as $\tr(G^\pm(t))=\bra{0} X_{\mu \nu}
\exp(\pm i H t-0 t) X_{\mu \nu}^\dagger\ket{0}$, and coherent states associated
to these. In this representation, we would have ended up with a Gaussian path
integral for the matrix Hamiltonian $H=\{H_{\mu \nu,\mu' \nu'}\}$ (first
quantization). However, we have not succeeded to extract the evolution along
approximately independent trajectories in $A$ and $B$ space at early times from
this representation.)  

Doing the Gaussian average over $H_A, H_B$ and $H_{AB}$, we arrive at a path integral with the action 
\begin{align*}
    S[\psi]&= \sum_{x=a,b}   \bar x \tau_3 (i \partial_t + i 0 \tau_3)
    x + \frac{i\lambda^2 N_X}{2} \tr(\Xi_X^2) \cr 
    &\qquad + \frac{i\Lambda^2 N_A N_B}{2}
     \tr\left( ((\Xi_A \odot \Xi_B)\tau_3)^2\right),
  \end{align*}  
where     $(\Xi_A)^{s s'}_{t t'} = -i N_A^{-1} a^s_{\mu,t} \bar a^{s'}_{\mu,t'}
(-)^{s'}$, $(\Xi_B)^{s s'}_{t t'} = -i N_B^{-1} b^s_{\nu,t} \bar b^{s'}_{\nu,t'}
(-)^{s'}$. Here, the traces include summation over all indices, i.e.
$\tr(XY)=\int dt dt'\sum_{s s'} X^{s s'}_{t t' }Y^{s' s}_{t' t}$, and similarly $\bar x X x\equiv \int dt dt' \,\bar x_tX_{t t'} x_{t'}$. The
explicit representation of the Hadamard product reads as $((\Xi_A \odot
\Xi_B)\tau_3)^{s s'}_{t t'}= \Xi^{s s'}_{ At t'}\Xi^{s s'}_{B t t'}(-)^{s'}$.
The dependence of the action on bilinears summed over Hilbert space indices
suggests the introduction of collective variables, $\Xi \equiv G $ via a
Lagrange multiplier locking: $\delta(\Xi_X-G_X)=\int D \Sigma_X\,\exp(- \tr(N_X
\Sigma_X(G_X-\Xi_X))$. Noting that $\tr(N_X \Sigma_X \Xi_X )=\exp( i\bar x
\Sigma_X \tau_3 x)$. With the free action given by $\sum_x \bar x \tau_3 (i \partial_t + i 0 \tau_3-\Sigma_X)
x $, the Gaussian integration over $x=a,b$ produces  the `$G \Sigma$' action (1) in the main text structurally 
similar to the action describing the SYK model \cite{rosenhausIntroductionSYKModel2019}. 

\subsection{Stationary phase action}

We consider the stationary phase ansatz $\bar G \to T \bar G T^{-1}\approx -
\frac{i}{\lambda} T \tau_3 T^{-1}$, where we used that, for slowly varying
matrices $T$, $(T \bar G T^{-1})_{t t'}=\int d\tilde t T_{t u}\bar G_{u-u'}
T^{-1}_{u' t'}\approx -\frac{i}{\lambda}(T \tau_3 T^{-1})_{t t'}\equiv
-\frac{i}{\lambda}Q_{t t'}$, on account of the temporal short-rangedness of
$\bar G$. Likewise, $\Sigma = -i \lambda Q $. Entering with this representation
into the action, the quadratic contributions $\tr(G_X^2) $ and $\tr(G \Sigma)$
decouple from the Goldstone modes (because $Q^2 =\mathds{1}$), while the
expansion of the `${\rm tr} \ln$' in $\partial_t$ acting on the Goldstone modes yields 
\begin{align*}
    &i N_X \tr\ln(i\partial_t + i \lambda Q)=i N_X \tr \ln (i \partial_t + i \lambda \tau_3 +i T^{-1}[\partial_t,T])\cr 
    &\qquad \approx i \frac{N_X}{\lambda}\tr(\tau_3 T^{-1}[\partial_t,T]). 
\end{align*}         
Noting that $N_X /\lambda = \pi \rho_X$ determines the spectral density of the
factor systems at the band center, and turning to an explicit integral
representation of the trace, we obtain Eq. (2) in the main text.

\subsection{$N_X^{-1}$-expansion}

The effect of the Hadamard vertex is best studied within in the framework of a
perturbative expansion of the Goldstone modes in generators, $T=\exp(W)$, $W=B
\tau_+ - B^\dagger \tau_-$, which, in view of the proportionality of the action
to $N_X$ is an $N_X^{-1}$ expansion. Considering its structure in causal space,
$\tr((G_A\odot G_B)^2) = \sum_{s s'}\tr(G_A\odot G_B)^{s s'}(G_A\odot G_B)^{s'
s}$, we realize that we get two structurally (and as we discuss in the main text
physically) different contribution of lowest, quartic order in $B$: in the terms
$s=s'$ of equal causality we expand one of the two factors $(G_A\odot G_B)^{s
s}\to \bar G^s_A\odot \bar G^s_B$ to zeroth order in the Goldstone mode
generators, and the other to quartic order, e.g., $(G_A\odot G_B)^{s s}\to -
(\lambda_A \lambda_B)^{-1} (B_A B_A^\dagger) \odot (B_B B_B^\dagger)$. We note
that the element-wise Hadamard product does not admit other terms at quartic
order. Substituting the time representation of $\bar G_X$, we arrive at the
first of Eqs. (4) in the main text. Turning to the causality mixing
channel,  $(G_A\odot G_B)^{+-}(G_A\odot G_B)^{-+ s}$, the expansion $G_X^{+-}\to
i \lambda_X B$ and $G_X^{-+}\to i \lambda_X B^\dagger$ immediately leads to the
second line of Eq. (4) in the main text. In the same way, we obtain the
Gaussian expansion of the free action stated above Eq. (3)
from the Goldstone mode action Eq. (2) in the main text. 
In the calculation of numerical coefficients of the coupling contants 
we assume that $B_{t,t'}$ is a slow function in $t+t'$ and independent of $t-t'$, which  
 e.g. gives rise to the Fermi's Golden Rule coupling rate 
$\gamma \approx\Lambda^2 /2 \lambda$
approximated by the  density of states at energy $\epsilon=0$.
A more accurate calculation accounting for all time-dependencies 
would rather lead to an energy-averaged density of states, but 
details are beyond the scope of this work.


\begin{thebibliography}{23}
\bibitem [{foo()}]{footnote1}%
  \BibitemOpen
  \href@noop {} {}\bibinfo {note} {While theory is based on approximations
  valid in the band center of the random matrix spectrum, the definition of the
  correlation function $K(t)$ in terms of an unrestricted trace over Green
  functions implies that the numerical analysis probes energies outside this
  region. This leads to quantitative uncertainties, which require the
  introduction of fitting constants.}\BibitemShut {Stop}%
\end{thebibliography}
\end{document}